\documentclass{INTERSPEECH2023}

\usepackage{multirow}
\usepackage{amssymb}
\usepackage{pifont}
\usepackage{hyperref}

\hypersetup{
    colorlinks=true,
    linkcolor=blue,
    filecolor=magenta,      
    urlcolor=cyan,
}

\newcommand{\cmark}{\ding{51}}%
\newcommand{\xmark}{\ding{55}}%

\interspeechcameraready

\title{PhonMatchNet: Phoneme-Guided Zero-Shot Keyword Spotting for User-Defined Keywords}
\name{Yong-Hyeok~Lee, Namhyun~Cho}
\address{
  Speech AI Lab., NCSOFT Corporation, South Korea} 
\email{\{eug92, cnh2769\}@ncsoft.com}

\begin{document}

\maketitle
 
\begin{abstract}
This study presents a novel zero-shot user-defined keyword spotting model that utilizes the audio-phoneme relationship of the keyword to improve performance. Unlike the previous approach that estimates at utterance level, we use both utterance and phoneme level information. Our proposed method comprises a two-stream speech encoder architecture, self-attention-based pattern extractor, and phoneme-level detection loss for high performance in various pronunciation environments. Based on experimental results, our proposed model outperforms the baseline model and achieves competitive performance compared with full-shot keyword spotting models. Our proposed model significantly improves the EER and AUC across all datasets, including familiar words, proper nouns, and indistinguishable pronunciations, with an average relative improvement of 67\% and 80\%, respectively. 
The implementation code of our proposed model is available at \href{https://github.com/ncsoft/PhonMatchNet}{https://github.com/ncsoft/PhonMatchNet}.

\end{abstract}

\noindent\textbf{Index Terms}: keyword spotting, user-defined, zero-shot, open-vocabulary

\section{Introduction}
Keyword spotting (KWS) in speech processing enables the identification of specific keywords in audio signals. Generally, conventional KWS methods require a significant amount of labeled data for training, which is a limitation in scenarios in which target keywords are rare, specialized, or user-defined \cite{huang22l_interspeech}. Recently, as user-friendly services, such as smart speakers, AI assistants, and personalized digital humans have become widespread, the demand for personalized technology that allows consumers to define their keywords instead of using pre-determined keywords set by manufacturers has increased. 
 
Zero-shot KWS is key to addressing this challenge, enabling a model to detect user-defined keywords with no prior training on those keywords. Most user-defined KWS (UDKWS) models are implemented using the query-by-example (QbyE) approach, which compares an input speech and a pre-registered speech in latent space \cite{chen2015query,lugosch2018donut,huang2021query}. Despite the success of the QbyE approach, it is limited in terms of performance and implementation because it compares the input speech with the pre-enrolled speech. 

Recently, a cross-modal correspondence detector (CMCD) that compares speech and text and can omit speech registration when adding new keywords has been proposed \cite{shin22_interspeech}. However, this method does not properly distinguish pairs of similar pronunciations, as the similarity between speech and text is calculated at the utterance level.

In this paper, we propose a novel zero-shot UDKWS model to address these issues. Our proposed model includes a two-stream speech encoder with a pre-trained speech embedder \cite{9053193}, self-attention-based pattern extractor \cite{li2020what}, and phoneme-level detection loss to achieve high performance in various pronunciation environments. We conducted experiments on datasets containing familiar words \cite{warden2018speech}, proper nouns \cite{kim2019query}, and indistinguishable pronunciations \cite{shin22_interspeech} to evaluate the effectiveness of our proposed model. Our proposed model outperforms the baseline model and achieves competitive performance compared with full-shot keyword spotting models. Particularly, our proposed model achieved a significant improvement in the equal error rate (EER) and area under the curve (AUC) across all datasets, with an average relative improvement of 67\% and 80\%, respectively.

\section{Related works}
This section examines recent UDKWS methods and related studies for our proposed method. The most recent model \cite{shin22_interspeech} showed degraded performance in similar pronunciations that are difficult to distinguish owing to the comparison of speech-text pairs at the utterance level. Furthermore, the audio encoder cannot properly handle uncommon pronunciations in the training dataset because it comprises only fully trainable modules. Thus, we employed a pre-trained speech embedder and phoneme-level detection loss to overcome these limitations.

Recently, various studies on non-semantic representation that allow a single embedder model to be used across various target tasks, not only in speech recognition \cite{baevski2020wav2vec} but also in other areas, such as keyword spotting \cite{9053193}, speaker identification, and language identification \cite{ShorJMLTQTSEH20_interspeech} have been presented. These studies improve models using techniques, such as self-supervised or unsupervised learning, multi-task learning \cite{PeplinskiSJGP21_interspeech}, and knowledge distillation \cite{ShorV22_interspeech}. Our objective is to obtain embeddings that can show high performance in general speech situations; therefore, we selected and froze the embedder model \cite{9053193} while considering the KWS performance and model size.

In various detection tasks using speech-text \cite{yusuf2021end}, visual-text \cite{PrajwalMAZ21_bmvc}, speech-visual \cite{MomeniASAZ20_bmvc}, and visual-only \cite{Prajwal22a}, temporal location information of targets significantly assists in achieving high performance. However, providing the correct labels for every timestamp in sequence data is challenging. Therefore, connectionist temporal classification (CTC) loss \cite{graves2006connectionist}, which does not require explicit alignment information, is widely used. However, unlike speech recognition in which speech and text labels always match, CTC loss is not appropriate for UDKWS, where speech and text labels may not match. Therefore, we propose a method that uses the sameness of speech and text pronunciation as a label rather than the exact time information.

\begin{figure*}[t]
    \centering
    \includegraphics[width=1.95\columnwidth]{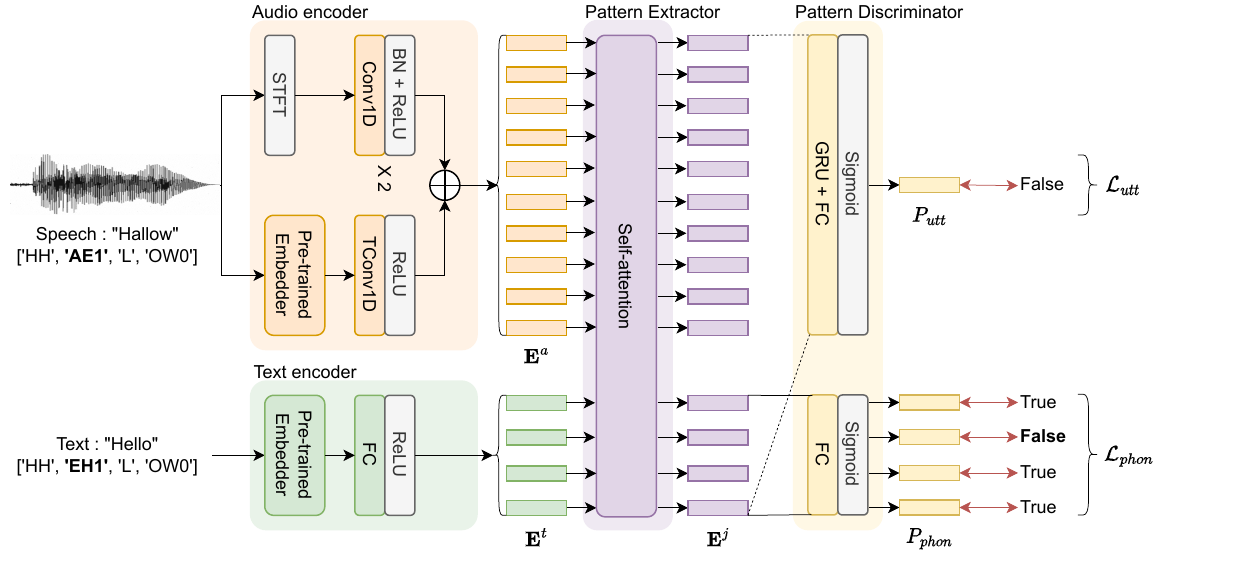}
    \caption{ Architecture of the proposed model. Boldface and red arrows denote the unmatched phonemes, labels, and BCE losses. ``TConv'' is the abbreviation of ``Transposed Convolution.''}
    \label{fig:overview}
\end{figure*}

\section{Proposed method}

Here, we describe our proposed model architecture and training criterion, as shown in Figure \ref{fig:overview}. Our proposed model comprises three sub-modules: an audio and text encoder with a pre-trained embedder, pattern extractor, and pattern discriminator. We use two loss functions based on the cross-entropy loss in the training criterion, which are computed at the utterance ($\mathcal{L}_{utt}$) and phoneme levels ($\mathcal{L}_{phon}$).

\subsection{Model architecture}

\noindent\textbf{Audio Encoder.} The audio encoder comprises two feature extractors: a pre-trained speech embedder \cite{9053193} with high performance in representing general pronunciations and a fully trainable feature extractor, which learns representations of special pronunciations, such as proper nouns. The pre-trained speech embedder has \SI{775}{\milli\second} windows and computes 96-dimensional feature vectors every \SI{80}{\milli\second}. We upsample the feature and time dimensions using a 1-D transposed convolution with a kernel size of 5 and stride of 4, along with fully connected layers. The fully trainable feature extractor comprises two 1-D convolutions with a kernel size of 3, batch normalization, and ReLU layers. The first convolution layer has a stride of 2, whereas the others have a stride of 1. The 40-dimensional mel-filterbank coefficients used as input are extracted every \SI{10}{\milli\second} with a window of \SI{25}{\milli\second}. Finally, we compute the 128-dimensional feature vectors extracted every \SI{20}{\milli\second} by adding the two feature vectors. We denote audio embeddings as $\mathbf{E}^a \in \mathbb{R}^{T_a\times128}$, where $T_a$ and 128 are the lengths of the audio and embedding dimension, respectively.

\noindent\textbf{Text Encoder.} The text encoder, similar to \cite{shin22_interspeech}, includes a pre-trained grapheme-to-phoneme (G2P) model\footnote{https://github.com/Kyubyong/g2p} followed by a fully connected layer and a ReLU activation function. We extract the G2P embedding from the last hidden states of the encoder. We denote text embeddings by $\mathbf{E}^t \in \mathbb{R}^{T_t\times128}$, which is the same as the audio encoder.


\noindent\textbf{Pattern extractor.} Considering the characteristics of the KWS task, which requires maintaining fewer model parameters, our pattern extractor is based on a self-attention \cite{li2020what, PrajwalMAZ21_bmvc} rather than a cross-attention mechanism \cite{MomeniASAZ20_bmvc, lee2021looking}. As in \cite{chefer2021generic}, during the fusion of multiple modalities, the self-attention method does not require other modules, unlike other attention mechanisms. The matrix of attention outputs for a set of queries $Q$ with keys and values packed into matrices $K$ and $V$ is computed as
\begin{equation} \label{eq:scaled_dot}
    \textrm{Attention}(Q,K,V)=\textrm{softmax}\left(\frac{QK^T}{\sqrt{d_k}}\right)V.
\end{equation}
Unimodal embeddings $\mathbf{E}^a$ and $\mathbf{E}^t$ are concatenated along the time dimension to prepare the joint embeddings $\mathbf{E}^j$:
\begin{equation} \label{eq:concat}
    \mathbf{E}^{c}=(\mathbf{E}^a;\mathbf{E}^t)\in\mathbb{R}^{(T_{a}+T_{t})\times128}.
\end{equation}
The concatenated embedding $\mathbf{E}^{c}$ is calculated as joint embeddings $\mathbf{E}^{j}$ using equation \ref{eq:scaled_dot}:
\begin{equation} \label{eq:attn_output}
    \mathbf{E}^{j}=\textrm{Attention}(\mathbf{E}^{c},\mathbf{E}^{c},\mathbf{E}^{c})\in\mathbb{R}^{(T_{a}+T_{t})\times128}.
\end{equation}
We used a lower triangular matrix as an attention mask to use only causal information of the intra-modality.

\noindent\textbf{Pattern discriminator.} Our pattern discriminator determines two probabilities, the match probability of audio and keywords and the audio and phoneme. To detect an utterance-level matching, we use a single GRU layer with 128 dimensions, taking the joint embeddings $\mathbf{E}^{j}$ along the time dimension as an input. The output of the last hidden state is fed into a fully connected layer with a sigmoid function:
\begin{equation} \label{eq:prob_utt}
P_{utt}=\sigma(\mathbf{W}^u\cdot\textrm{GRU}(\mathbf{E}^{j})+\mathbf{b}^u)\in\mathbb{R}^{1\times1}.
\end{equation}
Similarly, we extract only the phoneme sequences from the sequence of joint embeddings $\mathbf{E}^{j}$ and feed this into the fully connected layer with a sigmoid function to detect phoneme-level matching:
\begin{equation} \label{eq:prob_phon}
P_{phon}=\sigma(\mathbf{W}^p\cdot\mathbf{E}^{j}_\mathbf{s}+\mathbf{b}^p)\in\mathbb{R}^{T_t\times1},
\end{equation}
where $\mathbf{W}$, $\mathbf{b}$, $\sigma$, and $\mathbf{s}$ denote the trainable weights and biases of each fully connected layer, the sigmoid function, and the frame index of $(T_a,T_a+T_t]$, respectively.

\begin{table*}[t]
\centering
\caption{Performance of the baseline, proposed method, and ablations on various datasets. $\dagger$ refers to our implementation of H.-K. Shin \textit{et al.} $\mathbf{G}$, $\mathbf{Q}$, $\mathbf{LP_E}$, and $\mathbf{LP_H}$ refer to our test sets from Section \ref{section:datasets}. $\textbf{w/o Phoneme-level Detection Loss}$ represents the case of training using only utterance-level detection loss. $\textbf{w/o Self-attention-based Pattern Extractor}$ means the pattern extractor is composed of cross-modality attention rather than self-attention. $\textbf{w/o Fully trainable Speech Encoder}$ means the audio encoder comprises only a pre-trained speech embedder without convolution layers. We adjust the model parameters of the ablation studies to be the same as the proposed model.}
\label{tab:result}
\begin{tabular}{l|c|cccc|cccc}
\hline
\multicolumn{1}{c|}{\multirow{2}{*}{Method}} & \multirow{2}{*}{Params} & \multicolumn{4}{c|}{EER (\%) $\downarrow$}                                                                                                     & \multicolumn{4}{c}{AUC (\%) $\uparrow$}                                                                                                        \\ \cline{3-10} 
\multicolumn{1}{c|}{}                        &                             & \multicolumn{1}{c|}{$\mathbf{G}$}  & \multicolumn{1}{c|}{$\mathbf{Q}$}  & \multicolumn{1}{c|}{$\mathbf{LP_E}$} & $\mathbf{LP_H}$ & \multicolumn{1}{c|}{$\mathbf{G}$}   & \multicolumn{1}{c|}{$\mathbf{Q}$}   & \multicolumn{1}{c|}{$\mathbf{LP_E}$} & $\mathbf{LP_H}$ \\ \hline
$^{\dagger}$ CMCD \cite{shin22_interspeech}  & 653K                         & \multicolumn{1}{c|}{31.94}         & \multicolumn{1}{c|}{26.09}         & \multicolumn{1}{c|}{10.48}           & 29.34           & \multicolumn{1}{c|}{73.86}          & \multicolumn{1}{c|}{82.42}          & \multicolumn{1}{c|}{95.63}           & 77.60           \\
\textbf{Proposed}                            & 655K                         & \multicolumn{1}{c|}{\textbf{6.77}} & \multicolumn{1}{c|}{\textbf{4.75}} & \multicolumn{1}{c|}{\textbf{2.80}}   & \textbf{18.82}  & \multicolumn{1}{c|}{\textbf{98.11}} & \multicolumn{1}{c|}{\textbf{98.90}} & \multicolumn{1}{c|}{99.29}  & \textbf{88.52}  \\ \hline
{\quad}w/o Phoneme-level Detection Loss               & \multirow{3}{*}{655K}        & \multicolumn{1}{c|}{7.52}          & \multicolumn{1}{c|}{6.05}          & \multicolumn{1}{c|}{3.39}            & 19.65           & \multicolumn{1}{c|}{97.45}          & \multicolumn{1}{c|}{98.45}          & \multicolumn{1}{c|}{\textbf{99.30}}           & 88.04           \\
{\quad\quad}w/o Self-attention-based Pattern Extractor     &                             & \multicolumn{1}{c|}{11.32}         & \multicolumn{1}{c|}{18.24}         & \multicolumn{1}{c|}{3.47}            & 20.51           & \multicolumn{1}{c|}{95.73}          & \multicolumn{1}{c|}{90.14}          & \multicolumn{1}{c|}{99.26}           & 87.49           \\
{\quad\quad\quad}w/o Fully trainable Speech Encoder                       &                          & \multicolumn{1}{c|}{14.34}         & \multicolumn{1}{c|}{22.90}         & \multicolumn{1}{c|}{5.13}            & 25.77           & \multicolumn{1}{c|}{93.91}          & \multicolumn{1}{c|}{84.19}          & \multicolumn{1}{c|}{98.71}           & 82.53          \\ \hline

\end{tabular}
\end{table*}

\subsection{Training criterion}

Our training criterion ($\mathcal{L}_{total}$) comprises two binary-cross entropy (BCE) losses: an utterance- ($\mathcal{L}_{utt}$) and a phoneme-level detection losses ($\mathcal{L}_{phon}$),
\begin{equation} \label{eq:total_loss}
\mathcal{L}_{total}=\mathcal{L}_{utt}+\mathcal{L}_{phon}.
\end{equation}

\noindent\textbf{Utterance-level detection loss.} Similar to general KWS methods \cite{lopez2021deep}, the utterance-level detection loss is used to train the similarity between speech and keywords within a sample. The sample-level ground truth is used with $P_{utt}$ from Equation \ref{eq:prob_utt} to calculate the BCE loss.

\noindent\textbf{Phoneme-level detection loss.} We propose the phoneme-level detection loss to improve the performance of distinguishing similar pronunciations (e.g., ``friend'' and ``trend'') without using speech and pronunciation alignment information. The phoneme-level ground truth $\mathbf{y}$ is defined as 1 if the phoneme sequence of the speech label is the same as that of the keyword label and 0, otherwise.
\begin{equation} \label{eq:phon_loss}
  y_p = \begin{cases}
     1 & \text{if $y^t_p=y^s_p$} \\
     0 & \text{otherwise}
  \end{cases},
\end{equation}
where $y^s_p$, $y^t_p$, and $p$ denote the speech, keyword label, and phoneme index from 1 to $T_t$. As with $\mathcal{L}_{utt}$, the phoneme-level ground truth $\mathbf{y}$ and prediction $P_{phon}$ are used to calculate the BCE loss.

\section{Experiments}

Here, we describe the experimental setup, including the datasets, evaluation metrics, and implementation details for training and testing. Although most of our experimental settings are similar to those in \cite{shin22_interspeech}, we employ different methods to create anchor-negative pairs in certain test sets. An example of this is listed in Table \ref{tab:dataset} for clarity.

\begin{table}[t]
\caption{Examples of anchor and negatives of each dataset}
\label{tab:dataset}
\centering
\begin{tabular}{ccc}
\hline
\textbf{Dataset} & \textbf{Anchor} & \multicolumn{1}{c}{\textbf{Negatives}}                                                        \\ \hline
$\mathbf{G}$                & go              & \begin{tabular}[c]{@{}c@{}}yes\\ no\\ up\\ down\\ left\\ right\\ on\\ off\\ stop\end{tabular} \\ \hline
$\mathbf{Q}$                & hi galaxy       & \begin{tabular}[c]{@{}c@{}}hey android\\ hey snapdragon\\ hi lumina\end{tabular}              \\ \hline
\end{tabular}
\end{table}

\subsection{Datasets and metrics}
\label{section:datasets}

We used three datasets: LibriPhrase \cite{shin22_interspeech}, Google Speech Commands V1 ($\mathbf{G}$) \cite{warden2018speech}, and Qualcomm Keyword Speech dataset ($\mathbf{Q}$) \cite{kim2019query}, for training and evaluation. In the training phase, we used the training set of LibriPhrase and babble noise from the MS-SNSD dataset \cite{reddy2019scalable} for robust detection. Detailed training conditions were similar to that of \cite{shin22_interspeech}.
During evaluation, we used the datasets $\mathbf{G}$, $\mathbf{Q}$, LibriPhrase-easy ($\mathbf{LP_E}$), and LibriPhrase-hard ($\mathbf{LP_H}$). $\mathbf{LP_E}$ and $\mathbf{LP_H}$ are datasets reclassified as easy and difficult to differentiate between anchor and negative pairs in the test sets of LibriPhrase \cite{shin22_interspeech}. We evaluated our proposed models by measuring the EER and AUC at the sample level on these datasets. Because the datasets $\mathbf{G}$ and $\mathbf{Q}$ do not provide negative pairs, we calculated the EER and AUC by considering all keywords except the positive pairs as negative pairs among the candidate keywords for each dataset. Table \ref{tab:dataset} provides examples of anchor and negatives in $\mathbf{G}$ and $\mathbf{Q}$.

\subsection{Implementation details}

Our implementation was based on the TensorFlow library. The training criterion was optimized using the Adam optimizer with the default parameters. The models were trained for 100 epochs with a fixed learning rate of $10^{-3}$, and the best model was selected based on performance on the test sets. For training, we used a single V100 with a batch size of 2048 for a day.

\section{Results}
\subsection{Comparison with baseline}

We re-implemented the baseline model, CMCD, to compare its performance with that of our proposed model. According to Table \ref{tab:result}, our reconstructed CMCD performed similarly in $\mathbf{LP_E}$ and $\mathbf{LP_H}$ but showed decreased performance in $\mathbf{G}$ and $\mathbf{Q}$ datasets compared with the results of the original study \cite{shin22_interspeech}. This can be attributed to the difference in the method of generating anchor-negative pairs. Although the method differs from \cite{shin22_interspeech}, our method can measure the performance of these models with increased accuracy because it computes one anchor and other keywords. Our proposed model outperformed the baseline by a significant margin in all datasets, including familiar words (\textit{e.g.}, ``yes'' and ``no''), proper nouns (\textit{e.g.}, ``snapdragon'' and ``lumina''), and indistinguishable pronunciations (\textit{e.g.}, ``friend'' and ``frind''). A relative improvement of 67\% in EER and 80\% in AUC was observed on average across all the datasets.

\subsection{Ablation Study}

We evaluated the effectiveness of our proposed methods as shown in Table \ref{tab:result}: phoneme-level detection loss, self-attention-based modality fusion, and two-stream speech encoder.

\noindent\textbf{Effectiveness of Speech Encoder Components.} Our speech encoder comprised fully trainable and pre-trained modules. Comparing the first and last rows of Table \ref{tab:result}, we observed a considerable improvement in EER and AUC when using only the pre-trained speech embedder, except for the $\mathbf{Q}$ dataset. This could be attributed to the limitation that the embedder model we used was trained on general conversations from YouTube \cite{9053193}. Therefore, we compensated for this by adding fully trainable modules, which improved the performance on all test sets, including the $\mathbf{Q}$ dataset, as shown in the fourth row.

\noindent\textbf{Fusion Strategy for Audio-Text Representations.} Comparing the third and fourth rows of Table \ref{tab:result}, the performance of cross-modality attention and that of self-attention-based pattern extractor differed. Unlike the aforementioned analysis, we observed performance improvements on all test sets, particularly on the $\mathbf{Q}$ dataset.
The keywords of $\mathbf{Q}$ comprise proper nouns that are uncommon in typical training datasets. Unlike cross-modality attention-based models, a self-attention-based pattern extractor concatenated bi-modal data to produce outputs that contain all information in speech-to-speech, text-to-speech, and text-to-text modalities. This permitted the attention outputs, which are enhanced with speech information, to help distinguish the pronunciation similarity of rare cases in the training dataset.

\begin{figure}[t]
    \centering
    \includegraphics[width=0.9\columnwidth]{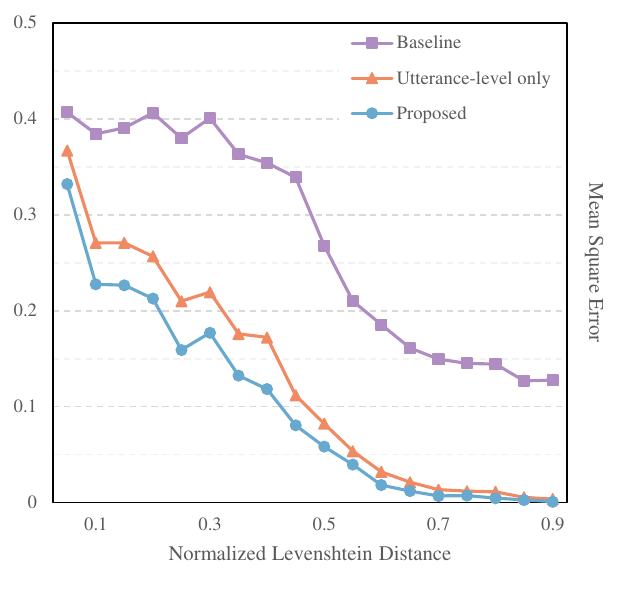}
    \caption{Evaluation results according to the normalized Levenshtein distance in a LibriPhrase test set. The smaller the distance, the more similar the pronunciation.}
    \label{fig:mse_ld}
\end{figure}

\noindent\textbf{Effectiveness of the Phoneme-level Detection Loss.} We proposed the phoneme-level detection loss to improve discrimination performance in speech-keyword pairs with duplicate pronunciations. As shown in Table \ref{tab:result} and Figure \ref{fig:mse_ld}, our proposed loss function outperformed the utterance-level detection loss only, particularly on the $\mathbf{LP_H}$ dataset and with low normalized Levenshtein distance, indicating the effectiveness of our proposed loss in discriminating speech-keyword pairs with similar pronunciations.

\subsection{Performance analysis via keyword similarity}

As shown in Figure \ref{fig:mse_ld}, we examined the relationship between phonetic similarity and model performance. We defined the similarity between speech-text label pairs using the normalized Levenshtein distance \cite{levenshtein1966binary} and calculated the mean squared error (MSE) between our model predictions and the labels over the similarity range greater than 0 and less than 1. Comparing our proposed model to the baseline, we observed superior performances across all distances. Notably, the error of the baseline model remained constant for normalized Levenshtein distances below 0.45, whereas our proposed model exhibited a linear improvement. Furthermore, we can attribute the performance improvements in similar pronunciations to the effectiveness of our phoneme-level detection loss.

\begin{table}[t]
\centering
\caption{Comparison of conventional KWS models and the proposed model in the test set of Google Speech Commands V1-12.}
\label{tab:conventional}
\begin{tabular}{lcc}
\hline
Model                                      & \multicolumn{1}{l}{Zero-shot} & Accuracy $\uparrow$      \\ \hline
DS-CNN \cite{9053395}               & \multirow{6}{*}{\xmark}       & 95.4           \\
Att-RNN \cite{seo2021wav2kws}                &                               & 95.6           \\
TC-ResNet \cite{choi2019temporal}          &                               & 96.6           \\
MHAtt-RNN \cite{rybakov2020streaming}      &                               & 97.2           \\
MatchBoxNet \cite{majumdar2020matchboxnet} &                               & 97.5 \\
KWT-3 \cite{berg2021keyword}               &                               & 97.5 \\
BC-ResNet-8 \cite{kim2021broadcasted}      &                               & \textbf{98.0}           \\ \hline
\textbf{Proposed}                                   & \cmark                        & \textbf{96.8}          \\ \hline
\end{tabular}
\end{table}


\subsection{Comparison with Conventional Methods}

We compared the performance of our proposed zero-shot KWS model with conventional full-shot KWS models in Table \ref{tab:conventional}. Because our model only calculated the matching probability, we set a threshold of 0.8 to compute the accuracy. Our proposed model outperformed some conventional models and achieved competitive performance compared with the state-of-the-art model. This result is noteworthy considering that we did not include the Google Speech Command dataset when training our model.

\section {Conclusions}

This study presented a novel user-defined keyword spotting model that leveraged the relationship between audio and phonemes of keywords. We introduced a two-stream speech encoder containing pre-trained embedders, a self-attention-based pattern extractor, and a phoneme-level detection loss to improve the performance of distinguishing similar pronunciations. Our proposed model outperformed the baseline model and achieved competitive performance compared with full-shot KWS models. However, despite our various efforts, relative performance improvements in challenging cases, such as $\mathbf{LP_H}$ were lower compared with other test sets. Improving the performance of highly similar pronunciations is our future research direction.
\\

\bibliographystyle{IEEEtran}
\bibliography{mybib}

\end{document}